# Flexible thermoelectrics in crossed graphene/hBN composites

M.Amir Bazrafshan, Farhad Khoeini*

*Department of Physics, University of Zanjan, P.O. Box 45195-313, Zanjan, Iran*

**Abstract**

Nanostructures exhibit unusual properties due to the dominance of quantum mechanical effects. In addition, the geometry of a nanostructure can have a strong influence on its physical properties. Using the tight-binding (TB) and force-constant (FC) approaches with the help of the non-equilibrium Green's function (NEGF) method, the transport and thermoelectric properties of cross-shaped (X-shaped) composite heterostructures are studied in two cases: Mixed graphene and h-BN (HETX-CBN) and all graphene (HETX-C) cross-shaped structures. Our numerical results show that an X-shaped structure helps to manipulate its electronic and phononic properties. The transport energy gap can be tuned in the range of ~0.8 eV by changing one arm width. Due to the drastic decrease in the electronic conductance of HETX-CBN and the dominance of the phononic thermal conductance, the ZT performance is degraded despite the high S value (in the order of meV). However, HETX-C has better ZT performance due to better electronic conductance and lower phononic/electronic thermal ratio, it can enhance the ZT ~2.5 times compared to that of zigzag graphene nanoribbon. The thermoelectric properties of the system can be tuned by controlling the size of the arms of the device and the type of its atoms.

**Keywords:** Graphene, Boron nitride, Heterostructure, Green's function, Tight-binding, X-shaped geometry, Force-constant.

## 1. Introduction

One of the active areas of nanoscience is 2D materials. Unconventional behavior is observed in the low dimensions due to quantum mechanical effects [1,2]. Hicks et al. [3] proposed that low-dimensional materials can provide high thermoelectric efficiencies (>1), giving hope after decades of efforts to achieve higher thermoelectric efficiency in the bulk form of materials. The advent of graphene has had a major impact on this field [4,5]. As the demand for energy increases, researchers are trying to address the impending energy crisis. One way to recycle waste heat, especially at the nanoscale, is through thermoelectrics [6]. Thermoelectric materials can be engineered at the nanoscale to crawl inside many electronic devices and recover the waste heat.

Hexagonal boron nitride is a wide bandgap semiconductor [7–11], while graphene is a semimetal [5]. The study of the properties of carbon-based materials that exhibit different behaviors is also an active field [12–18]. The shape of a nanostructure is important in determining its physical properties [19,20]. As a well-known example, the graphene nanoribbon is not necessarily a zero-gap semiconductor. Armchair graphene nanoribbons (AGNRs) can exhibit two different electronic behaviors, from a metal to an insulator, depending on their width [21]. While the behavior of AGNRs is width dependent, zigzag graphene nanoribbons (ZGNRs) are always metallic in the simple tight-binding approach, regardless of width. However, all hBN nanoribbons are semiconductors [22,23], which by a certain edge passivation, zigzag hBN nanoribbons can turn into a metal [8]. There is a small lattice mismatch between graphene and hBN [24,25], meaning they can couple to each other. However, the in-plane carbon-boron-nitride heterojunction has been realized experimentally [26].

The thermoelectric performance can be evaluated by $ZT(\mu, T) = T \frac{G(\mu,T)S^2(\mu,T)}{\kappa_e(\mu,T)+\kappa_{ph}(T)}$, where T is the temperature, G is the electronic conductance (which is a function of the chemical potential µ and T), S is the Seebeck coefficient, $\kappa_e$ is the electronic thermal conductance, and $\kappa_{ph}$ is the phononic thermal conductance. From a thermoelectric point of view, the larger the temperature difference between two points means the better energy difference and hence the better thermoelectric performance. One may conclude that insulators are generally better candidates for this purpose. In insulators, electrons cannot contribute to heat transport. But there should be a limit, if electrons cannot move at all, then there is no electric current, and the whole story of converting heat into electricity becomes meaningless.

Using the quantum mechanical description provided by the tight-binding [27] and the force-constant methods implemented in the non-equilibrium Green's function formalism [28], the electronic and phononic transport properties of the cross-shaped heterojunction composed of graphene and hBN are investigated. As mentioned earlier, geometry is important at the nanoscale, and here we are interested in studying a cross-shaped geometry of a planar carbon boron nitride heterostructure.

Curvatures can alter the transport behavior of a nanostructure [29–31]. However, the available energy levels in the device section of a transport configuration are important [32]. Geometry, doping, vacancies, introducing strain [33,34], and applying an electric (or magnetic) field are all ways to change these levels, and thus changing the transport coefficient. Other electronic transport properties, e.g. the Seebeck coefficient, electronic thermal conductivity, current through the structure, and electronic conductance, are directly dependent on this quantity. In HETX-CBN, one arm is an armchair hBN nanoribbon with a width of 9 atoms (9-ABNNR), which is passed through the other arm, an N-AGNR with varying widths (N=8-10). Thus, in the device part of the transport system we have at least three different nanoribbons, a 9-ABNNR, an N-AGNR, and a zigzag graphene nanoribbon with a width of 38 atoms. Mixing these different systems in an X-shaped structure can lead to unexpected energy level configurations. The presence of two crossing arms can increase the tunability of electronic and phononic properties, as suggested by numerical results. By changing the width of the arms, we expect the electronic and phononic properties to change, which is confirmed by the numerical results. The contribution of high-frequency phonons to thermal conductance increases at high temperatures. Tailoring the structure into an X-shape blocks high-frequency phonons. It should be noted that the all-carbon X-shaped structure is not a heterostructure, but for convenience and to maintain simple notation in the naming convention, we preserve the letters "HET" for it.

In the following section, we describe the model and the methodology. The results are discussed in the third section. The conclusion is presented in the last section.

## 2. Model and Method

In this section we describe the structures studied and briefly present the computational method.
We choose crossed structures in which one of its arms is an armchair nanoribbon with a width of 9 atoms, i.e. this is our fixed arm and it doesn't change in all structures (Figure 1, blue shaded arm). The fixed arm is the one that connects the source from the bottom left to the top right in Figure 1. The fixed arm is assumed to pass through the other arm, so the variable-width arm has two distinct parts. We studied the width variation of the variable-width AGNR arms (from 8 to 10 atoms in width, pale-colored atoms in Figure 1) for different configurations composed of graphene hBN HETX and fully graphene HETX, see Figure 1 (a) and (b), respectively. The naming convention is based on the arms elements, and two parts of the variable width arm, when the elements of the arms are only carbon atoms, and the upper part of the variable arm width has the width of 8 atoms (arrow indicated by $W_{A1}$ in Figure 1) followed by the 9 of the lower part (arrow indicated by $W_{A2}$), we call it C-8-9.

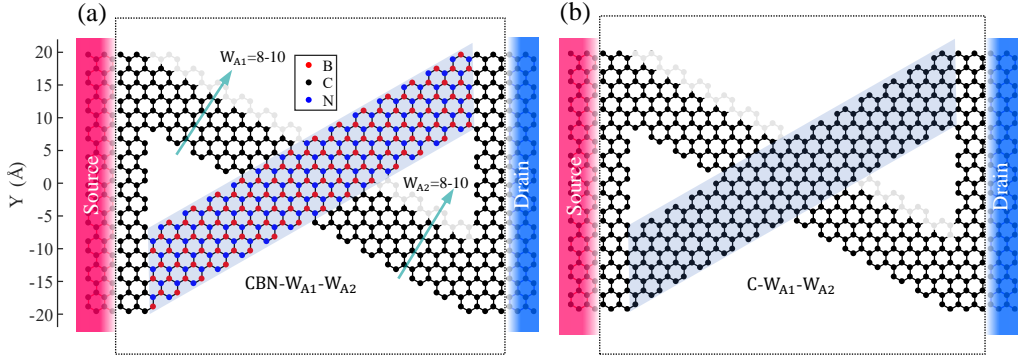

Figure 1. Schematic of the crossed heterostructures studied. (a) CBN-$W_{A1}$-$W_{A2}$ HETX, and (b) C-$W_{A1}$-$W_{A2}$ HETX. Pale atoms are related to the variation of the variable width arm. The blue shading indicates the fixed arm. The length of the device is ~5.8 nm and the device is indicated by a dotted rectangle.

The electronic properties are studied using the TB Hamiltonians implemented in the NEGF formalism. The Hamiltonians in the TB approximation are given in [35,36]:

$$H = \sum_i \varepsilon_i |i\rangle\langle i| + \sum_{\langle i,j \rangle} t_{i,j}|i\rangle\langle j| + \text{H. c.}, \quad (2)$$

where $\varepsilon_i$ is the on-site energy and $t_{i,j}$ is the hopping parameter between atoms $i$ and $j$. The TB parameters for carbon atoms are taken from [37], and for hBN from [22]. The hopping between carbon, boron and nitride is also evaluated by averaging the hopping parameters, as worked out in [38]. The TB parameters are listed in Table 1. Details of the calculation method are given in [39].

The Hamiltonians of the TB approach are then implemented in the NEGF formalism. The retarded Green's function is [40]:

$$G(E) = [(E + i\eta)I - H_C - \Sigma_{SC}(E) - \Sigma_{DC}(E)]^{-1}, \quad (3)$$

where $E$ is the electron energy, I is the identity matrix, $\eta$ is an arbitrarily small positive number, $H_C$ is the central scattering region (or device) Hamiltonian, and the self-energy for the source (drain) electrode is represented by $\Sigma_{SC(DC)}$. Details are given in [41].

The spectral density operator can be calculated as:

$$\Gamma_{S(D)}(E) = i[\Sigma_{SC(DC)}(E) - \Sigma_{SC(DC)}(E)^\dagger], \quad (4)$$

The electronic transmission probability (or the transport coefficient) can be obtained as:

$$T_e(E) = \text{Trace}[\Gamma_S(E)G(E)\Gamma_D(E)G(E)^\dagger]. \quad (5)$$

The electronic conductance (G), the Seebeck coefficient (S), and the electronic thermal conductance ($\kappa_e$) can be calculated as [31,42–44]:

$$G(\mu, T) = e^2 L_0(\mu, T), \quad (6)$$

$$S(\mu, T) = \frac{1}{eT}\frac{L_1(\mu, T)}{L_0(\mu, T)}, \quad (7)$$

$$\kappa_e(\mu, T) = \frac{1}{T}\left[L_2(\mu, T) - \frac{L_1^2(\mu, T)}{L_0(\mu, T)}\right], \tag{8}$$

where $e$ is the elementary charge and $L_n$ is given by:

$$L_n(\mu, T) = -\frac{2}{hk_BT}\int_{-\infty}^{\infty} T_e(E)(E-\mu)^n \frac{\exp(\frac{E-\mu}{k_BT})}{\left(\exp(\frac{E-\mu}{k_BT}) + 1\right)^2} dE, \tag{9}$$

with $h$ is the Plank constant and $k_B$ is the Boltzmann constant. The temperature is assumed to be 300 K in the calculations.

To study the phononic properties, we use the force-constant method with the interactions up to four nearest neighbors (4NN). The secular equation for phonons, which derived from Newton's second law, is [44]:

$$DU = \omega^2 U, \tag{10}$$

with $U$ is the matrix containing the vibrational amplitude of all atoms, $\omega$ is the angular frequency, and $D$ is the dynamical matrix given by:

$$D = [D_{i,j}^{3\times 3}] = \begin{cases} -\frac{K_{i,j}}{\sqrt{M_iM_j}} & \text{for } j \neq i \\ \sum_{j\neq i}\frac{K_{i,j}}{M_i} & \text{for } j = i \end{cases}. \tag{11}$$

In this relation, $M_i(M_j)$ is the mass of the $i^{th}(j^{th})$ atom, and $K_{i,j}$ represents the $3 \times 3$ force tensor between the $i^{th}$ and the $j^{th}$ atom as:

$$K_{i,j} = U^{-1}(\theta_{i,j})K_{i,j}^0 U(\theta_{i,j}), \tag{12}$$

with $\theta_{ij}$ is the angle between the $i^{th}$ and the $j^{th}$ atoms. The unitary matrix $U(\theta_{i,j})$ is defined by the rotation matrix in a plane as:

$$U(\theta_{i,j}) = \begin{pmatrix} \cos\theta_{i,j} & \sin\theta_{i,j} & 0 \\ -\sin\theta_{i,j} & \cos\theta_{i,j} & 0 \\ 0 & 0 & 1 \end{pmatrix}, \tag{13}$$

with a $K_{i,j}^0$ given by:

$$K_{i,j}^0 = \begin{pmatrix} \varphi_r & 0 & 0 \\ 0 & \varphi_{t_i} & 0 \\ 0 & 0 & \varphi_{t_o} \end{pmatrix}, \tag{14}$$

where $\varphi_r, \varphi_{t_i}, \varphi_{t_o}$ are force constant parameters in the radial, in-plane and out of plain directions of the $j^{th}$ atom, respectively. The important consideration in these matrices is for the $D_D$, which represents the dynamical matrix of the device section, regarding Eq. 12, and to write what each atom feels (or when $i = j$), one has to consider all 4NN effects in the summation, including atoms in the neighboring unit cells, i.e., $D_{D_{i,j}} = \sum_i \sum_{j \in 4NN} K_{iDevice,jDevice} + K_{iDevice,jSource} + K_{iDevice,jDrain}$. For coupling terms such as the elements of $D_{Device-Drain}$, the interaction between the first atom of the device and the first atom of the drain (i.e., diagonal elements) is already accounted, so one can safely set this to zero. The force constants are taken from [45–47] and are also given in Table 1.

The phononic transmission probability ($T_{ph}$) can be obtained by the Green's function method, using $\omega^2$ instead of E in Eq.3. The phononic thermal conductance is given by [48]:

$$\kappa_{ph}(T) = \frac{1}{8\pi k_B T^2} \int_0^\infty \hbar^2 \omega^2 \frac{T_{ph}(\omega)}{\sinh^2\left(\frac{\hbar\omega}{2k_B T}\right)} d\omega. \qquad (15)$$

The phononic transport can be considered ballistic when $L_D$ is much smaller than the phonon mean free path [49], but it is also a function of the width of the GNR [49,50]. In addition, the phononic band structure can be evaluated by solving the following eigenvalue problem:

$$\left(\sum_i \sum_j D_{i,j} - \omega^2(q)I\right)\delta_{i,j} - \sum_i \sum_j D_{i,j} e^{iq\cdot\Delta r_{i,j}} = 0. \qquad (16)$$

In this equation, $\Delta r_{i,j} = r_i - r_j$ is the distance between the $i^{th}$ and the $j^{th}$ atoms, and $q$ is the wave vector. Note that the first term forms the dynamical matrix of the device, i.e., $D_D$.

Table 1. The TB and FC parameters [22,37,38,45–47].

| Element | On-site energy (eV) | Interacting elements | Hopping energy (eV) |
|---|---|---|---|
| C | 0 | C-C | -2.7 |
| B | 2.315 | C-B | -2.55 |
| N | -2.315 | C-N | -2.55 |
| - | - | B-N | -2.4 |
| **Mass of the carbon atom ($M_C$)** | | | $1.994 \times 10^{-26}$ (Kg) |
| **Mass of the boron atom ($M_B$)** | | | $1.795 \times 10^{-26}$ (Kg) |
| **Mass of the nitrogen atom ($M_N$)** | | | $2.325 \times 10^{-26}$ (Kg) |

| Force-Constant parameters (N/m) | | | | | | |
|---|---|---|---|---|---|---|
| | C-C | B-B | N-N | B-C | N-C | B-N |
| $\varphi_r^1$ | 398.7 | 0 | 0 | 354.35 | 354.35 | 310 |
| $\varphi_{t_i}^1$ | 172.8 | 0 | 0 | 178.9 | 178.9 | 185 |
| $\varphi_{t_o}^1$ | 98.9 | 0 | 0 | 77.45 | 77.45 | 56 |
| $\varphi_r^2$ | 72.9 | 70 | 80 | 71.45 | 76.45 | 0 |
| $\varphi_{t_i}^2$ | -46.1 | -32.3 | -7.3 | -39.2 | -26.7 | 0 |
| $\varphi_{t_o}^2$ | -8.2 | -7 | -5.5 | -7.6 | -6.85 | 0 |
| $\varphi_r^3$ | -26.4 | 0 | 0 | -8.2 | -8.2 | 10 |
| $\varphi_{t_i}^3$ | 33.1 | 0 | 0 | 0.3 | 0.3 | -32.5 |
| $\varphi_{t_o}^3$ | 5.8 | 0 | 0 | 6.15 | 6.15 | 6.5 |
| $\varphi_r^4$ | 1 | 0 | 0 | -9 | -9 | -19 |
| $\varphi_{t_i}^4$ | 7.9 | 0 | 0 | 10.4 | 10.4 | 12.9 |
| $\varphi_{t_o}^4$ | -5.2 | 0 | 0 | -4.1 | -4.1 | -3 |

## 3. Results and discussion

The transmission spectrum is the key component in evaluating the electronic transport properties of a material. The electronic behavior of an AGNR is strongly width dependent [51]. In this work, we consider three widths of an arm (Figure 1, the blue shaded arm) in the HETX structures, ranging from 8 to 10 atoms in width. In the following, when we discuss the arms in the HETX structure, we mean the variable-width arm, not the fixed-width arm, represented by $W_{A1}$ and $W_{A2}$ in Figure 1.

The electronic and phononic band structures (dispersion plots) of different nanoribbons involved in all the structures are shown in Figure 2, including electrodes that are two semi-infinite 38-ZGNR (Figure 2 (a)), 8-AGNR (Figure 2 (b)), 9-AGNR (Figure 2 (c)), 10-AGNR (Figure 2 (d)), 9-ABNNR (Figure 2 (e)), as a different variable arms of the HETX structures. As can be seen, 8-AGNR is metallic, but 9 and 10-AGNR are semiconductors, while 9-ABNNR is a wide-gap semiconductor. Also, the phononic band structure of 9-ABNNR shows that the group velocity of phonons in 9-ABNNR is lower than that of graphene nanoribbons (GNRs). In addition, its frequency range is smaller than that of GNRs.

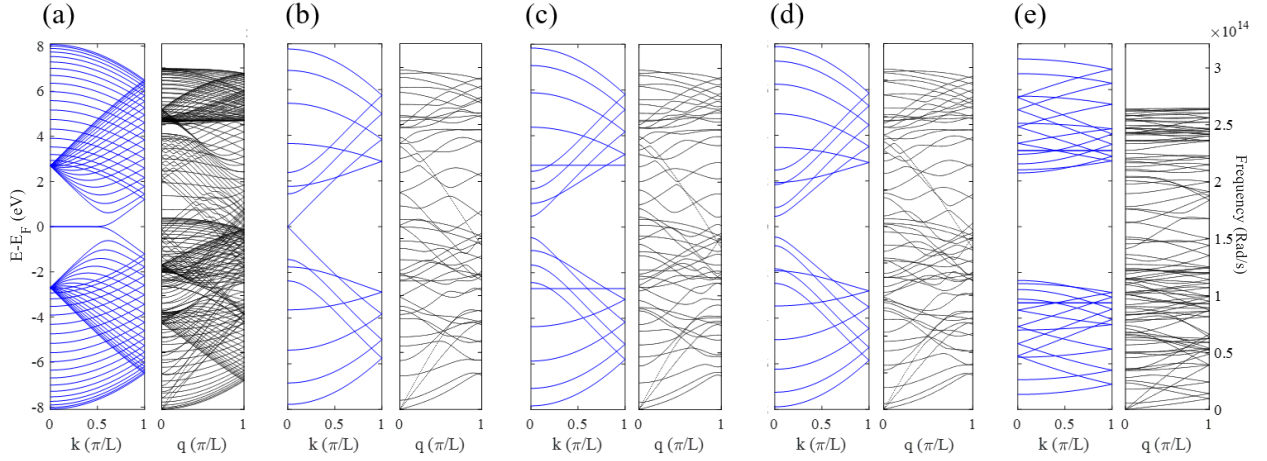

Figure 2. Electronic and phononic dispersions for all nanoribbons in the structures investigated, (a) 38-ZGNR, (b) 8-AGNR, (c) 9-AGNR, (d) 10-AGBR, and (e) 9-ABNNR.

The electronic and phononic transmission probabilities are then studied. As shown in the transmission plots in Figure 3, all HETX-CBNs are semiconducting and all HETX-Cs are metallic. The variation of the energy gap of the HETX-CBNs is shown in Table 2. The results show that changing the width of the arms in HETX-CBN can help to control the transport energy gap. The conducting states of ZGNRs (in the TB) are on the edge atoms, and the existence of a transmission value at the Fermi energy suggests that the ZGNR character is more pronounced. However, if one of the arms is removed, an S-shaped structure is obtained. As shown in [44], for an S-shaped structure with an AGNR as the device, the ZGNR metallic states are also present in the transmission spectrum, but adding an arm increases the transmission probability at other energies, showing a better coupling of the energy levels in the device and the electrodes. It is important for a good thermoelectric material to be good at conducting electricity while being thermally insulating [52].

Table 2. Transport energy gap for the HETX-CBN structures studied

| Configuration | CBN-8-8 | CBN-8-9 | CBN-9-9 | CBN-10-9 | CBN-10-10 |
|---|---|---|---|---|---|
| **Transport energy gap (eV)** | 1.2 | 1.42 | 1.75 | 1.86 | 1.95 |

In CBN-10-10, low frequency phonons were significantly reduced. In addition, HETX-CBN shows significant suppression in the transmission spectrum of high frequency phonons, indicating its better performance in reducing the thermal conductivity of the lattice due to the appearance of the ABNNR arm character.

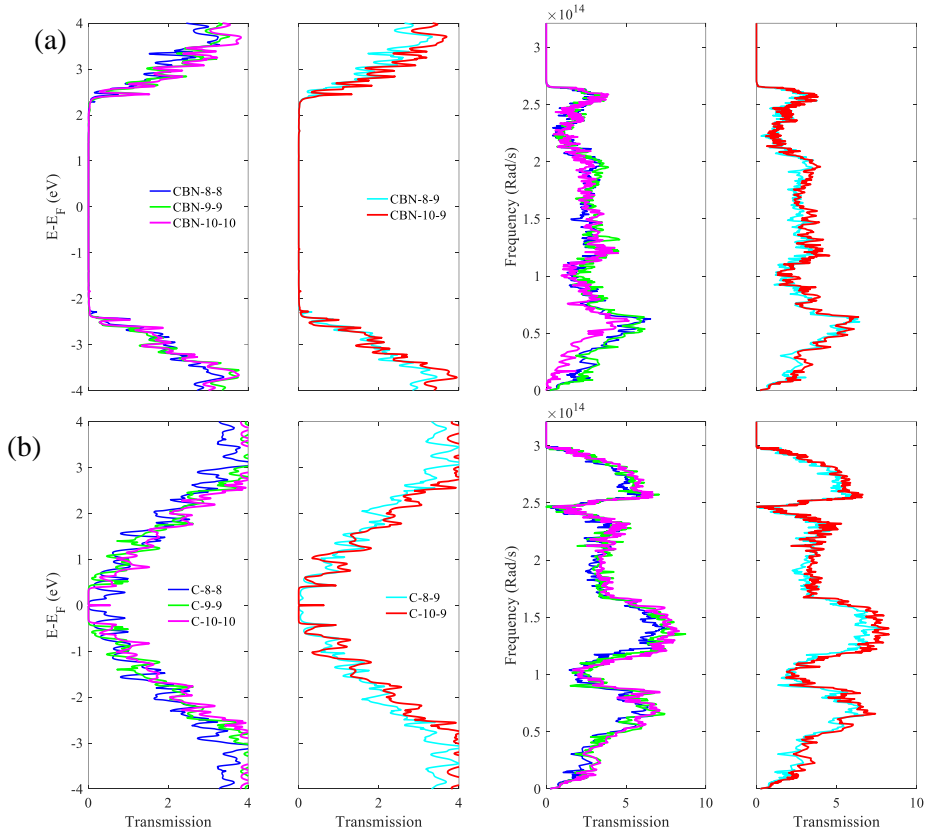

Figure 3. Electronic and phononic transmission spectra of HETXs for CBN-8-8 (blue), CBN-8-9 (cyan), CBN-9-9 (green), CBN-10-9 (red) and CBN-10-10 (magenta) (a). The same as shown in (a) but for HETX-C (b).

The thermoelectric related parameters are then examined. In Figure 4 (a), G, S, and $\kappa_e$ are plotted against the chemical potential at 300 K for HETX-CBN. Although the S in HETX-CBN reaches high values of 3 mV/K due to the large transport energy gap, the ZT value is suppressed due to the significant reduction in G. The Seebeck coefficient increases in the presence of an energy gap, but very small values of G strongly degrade the ZT, as suggested by the ZT formula. The electronic thermal conductance shows very small values of $\mathcal{O} \times 10^{-13}$ W/K, which is a consequence of the suppressed G.

However, the same parameters are studied for the HETX-C in Figure 4 (b). For the all-carbon X-shaped nanostructures, although the electronic conductance is not significantly suppressed in the chemical potential range of -1 to 1 compared to HETX-CBN, the absence of the transport gap strongly degrades the S. Moreover, the Seebeck coefficient responds to any energy gap, even those on one side of the energy spectrum with respect to the Fermi energy. However, to access these gaps one has to change the chemical potential, which in real situations can change the electronic structure. Therefore, it is of great importance for practical applications to obtain better values of these parameters in a chemical potential as close to the Fermi energy as possible. The last parameter in this figure, $\kappa_e$, has a value of $\mathcal{O} \times 10^{-10}$ W/K. As we noted earlier, G and $\kappa_e$ are related to each other, when G is reduced, the electronic thermal conductance is also reduced. However, in some cases it is possible to break this close behavior [44]. On the other hand, the electronic parameters should be tuned with respect to $\kappa_{ph}$, meaning the ratio of $\kappa_{ph}/\kappa_e$ is important. If this ratio is too large, the ZT will be significantly reduced due to the phononic thermal conductance.

The comparison of the electronic conductance between HETX-CBN (Figure 4 (a)) and HETX-C (Figure 4 (b)) shows that the electronic quantities are almost symmetric with respect to the Fermi energy for the case of all-carbon device, while for HETX-CBN there is no such a symmetry, showing that the hBN arm can play an important role in tuning the electronic properties and breaking the electron-hole symmetry. This can be attributed to the high ionicity of the B and N elements.

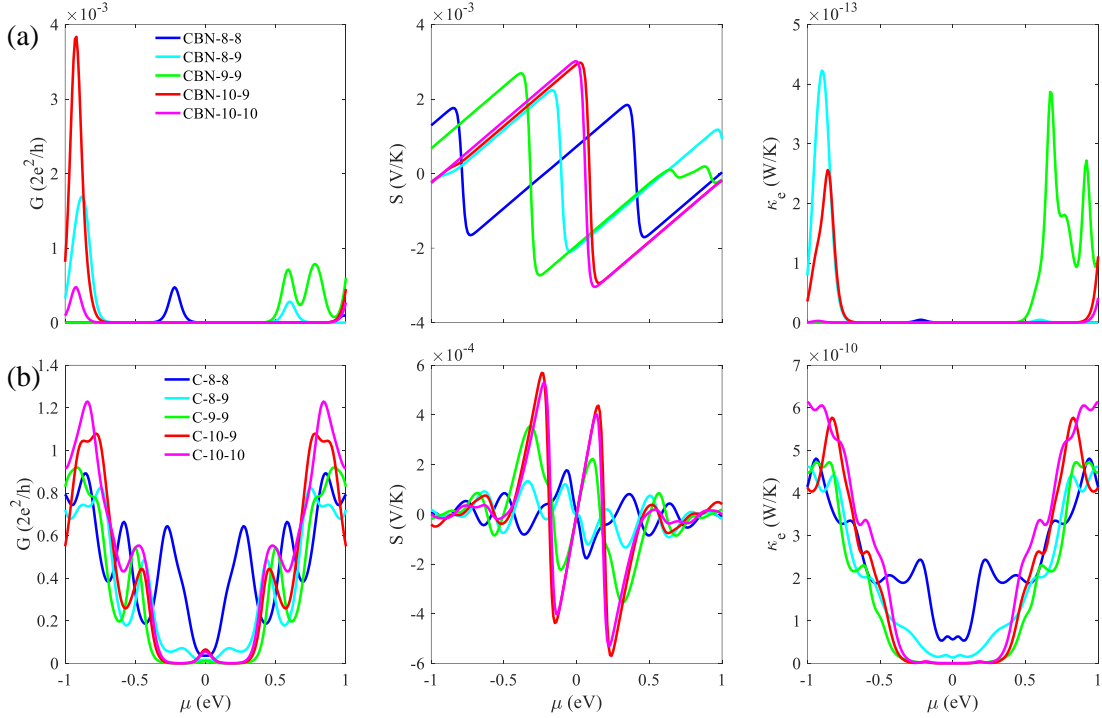

Figure 4. The G (left panel), S (middle panel), and $\kappa_e$ (right panel) of HETXs for CBN-8-8 (blue), CBN-8-9 (cyan), CBN-9-9 (green), CBN-10-9 (red), and CBN-10-10 (magenta) (a). The same as in (a) but for HETX-C (b).

The phononic thermal conductivity and ZT are then examined in Figure 5 by plotting $\kappa_{ph}$ as a function of temperature. As can be seen in the left panel of Figure 5 (a), the $\kappa_{ph}$ for CBN-10-10 is $\mathcal{O} \times 10^{-8}$ W/K (due to suppressed transmission at low frequencies), and the other HETX-CBN configurations are of $\mathcal{O} \times 10^{-9}$ W/K, which means that $\kappa_{ph}/\kappa_e$ is at least of $\mathcal{O} \times 10^4$. This large value implies that the phononic thermal conductance is dominant, so one can safely neglect $\kappa_e$. However, the $\kappa_{ph}/\kappa_e$ for HETX-Cs are of $\mathcal{O} \times 10^1$, see left panel of Figure 5 (b). This figure also shows that HETX-C does not significantly reduce the phononic thermal conductivity. Nevertheless, the configurations with an 8-AGNR arm suppress the $\kappa_{ph}$ more than other HETX-Cs, which is a consequence of the size. On this basis, neglecting phononic thermal conductivity requires special care even at the nanoscale and under the assumption of ballistic transport. As size decreases, the number of atoms in the structure decreases, and thus the number of available phononic modes decreases. In general, matching the phononic modes improves the thermal conductivity and vice versa.

The final part of this study is to examine the ZT. The right panel of Figure 5 shows the ZT as a function of the chemical potential for HETX-CBN. Although the S has large values, we expected a weak ZT performance because of the degraded G and the high $\kappa_{ph}/\kappa_e$ ratio. The right panel of Figure 5 (a) shows a ZT of $\mathcal{O} \times 10^{-3}$, confirming the expectation. The situation is better for HETX-C. Although the S is one order of magnitude smaller than that of HETX-CBN, the ZT performance is two orders of magnitude larger than that of HETX-CBN, reaching a value of 0.13 at $\mu = \pm 0.42$ eV. The ZT plots suggest that the various parameters of the ZT formula are closely related to significant changes in the transmission spectrum. This conclusion is valid at low temperatures. However, as the temperature increases, the sharp change in the transmission spectrum becomes smoother and the interpretation of the ZT formula becomes more complex.

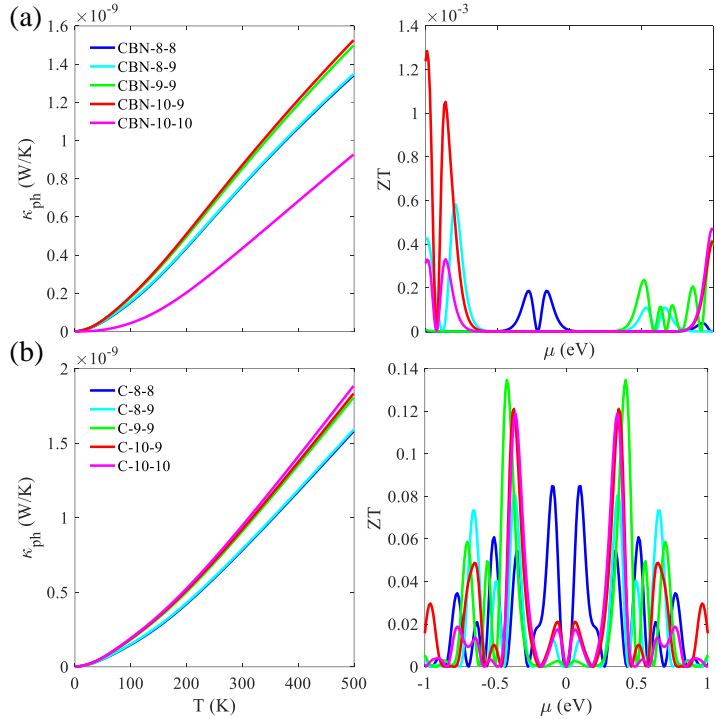

Figure 5. The $\kappa_{ph}$ as a function of temperature (left panel) and ZT vs. chemical potential (right panel) of HETXs for CBN-8-8 (blue), CBN-8-9 (cyan), CBN-9-9 (green), CBN-10-9 (red), and CBN-10-10 (magenta) (a). The same as shown in (a) but for HETX-C (b).

## 4. Conclusions

We have studied the electronic transport and thermoelectric properties of cross structures, the HETX-CBN and HETX-C, by considering one arm as a fixed arm and varying the width of the other arm from 8 to 10 atoms in width. The results show that the energy gap of HETX-CBNs can be strongly tuned by varying the variable arm width. The transport energy gap of HETX-CBN can be tuned by ~0.8 eV. In addition, CBN-10-10 can significantly change the phononic transport properties by suppressing the phononic transmission spectra at low frequencies. Moreover, the HETX-C structures are metallic and show a symmetric behavior with respect to the Fermi energy, while the HETX-CBN doesn't show such symmetric behavior.

Despite the high Seebeck coefficient of HETX-CBN in the order of meV, the ZT performance of HETX-C is significantly better than that of HETX-CBN due to the non-zero electronic conductance of the HETX-C structure in the chemical potential range investigated.

## Data availability

All data generated for this study are included in the manuscript.

## Author contributions

M.A.B carried out the simulations. M.A.B and F.K analyzed the data and prepared the manuscript. F.K. supervised the project and revised the final manuscript. All authors read and approved the final manuscript.

# Competing interests

The authors declare no competing interests.

*Email: khoeini@znu.ac.ir*